\documentclass[conference,a4paper]{IEEEtran}
\usepackage[dvips]{graphicx,color}
\usepackage{latexsym}
\usepackage{amssymb}
\usepackage{array}

\begin{document}

%
%
\newcommand{\qed}{\hfill$\square$}
\newcommand{\suchthat}{\mbox{~s.t.~}}
\newcommand{\markov}{\leftrightarrow}
\newenvironment{pRoof}{%
 \noindent{\em Proof.\ }}{%
 \hspace*{\fill}\qed \\
 \vspace{2ex}}


\newcommand{\ket}[1]{| #1 \rangle}
\newcommand{\bra}[1]{\langle #1 |}
\newcommand{\bol}[1]{\mathbf{#1}}
\newcommand{\rom}[1]{\mathrm{#1}}
\newcommand{\san}[1]{\mathsf{#1}}
\newcommand{\mymid}{:~}
\newcommand{\argmax}{\mathop{\rm argmax}\limits}
\newcommand{\argmin}{\mathop{\rm argmin}\limits}

\newcommand{\Cls}{class NL}
\newcommand{\vSpa}{\vspace{0.3mm}}
\newcommand{\Prmt}{\zeta}
\newcommand{\pj}{\omega_n}

\newfont{\bg}{cmr10 scaled \magstep4}
\newcommand{\bigzerol}{\smash{\hbox{\bg 0}}}
\newcommand{\bigzerou}{\smash{\lower1.7ex\hbox{\bg 0}}}
\newcommand{\nbn}{\frac{1}{n}}
\newcommand{\ra}{\rightarrow}
\newcommand{\la}{\leftarrow}
\newcommand{\ldo}{\ldots}
\newcommand{\typi}{A_{\epsilon }^{n}}
\newcommand{\bx}{\hspace*{\fill}$\Box$}
\newcommand{\pa}{\vert}
\newcommand{\ignore}[1]{}

%
%
%
%
\newcommand{\bc}{\begin{center}}  %
\newcommand{\ec}{\end{center}}
\newcommand{\befi}{\begin{figure}[h]}  %
\newcommand{\enfi}{\end{figure}}
\newcommand{\bsb}{\begin{shadebox}\begin{center}}   %
\newcommand{\esb}{\end{center}\end{shadebox}}
\newcommand{\bs}{\begin{screen}}     %
\newcommand{\es}{\end{screen}}
\newcommand{\bib}{\begin{itembox}}   %
\newcommand{\eib}{\end{itembox}}
\newcommand{\bit}{\begin{itemize}}   %
\newcommand{\eit}{\end{itemize}}
\newcommand{\defeq}{\stackrel{\triangle}{=}}
\newcommand{\Qed}{\hbox{\rule[-2pt]{3pt}{6pt}}}
\newcommand{\beq}{\begin{equation}}
\newcommand{\eeq}{\end{equation}}
\newcommand{\beqa}{\begin{eqnarray}}
\newcommand{\eeqa}{\end{eqnarray}}
\newcommand{\beqno}{\begin{eqnarray*}}
\newcommand{\eeqno}{\end{eqnarray*}}
\newcommand{\ba}{\begin{array}}
\newcommand{\ea}{\end{array}}
\newcommand{\vc}[1]{\mbox{\boldmath $#1$}}
\newcommand{\lvc}[1]{\mbox{\scriptsize \boldmath $#1$}}
\newcommand{\svc}[1]{\mbox{\scriptsize\boldmath $#1$}}

\newcommand{\wh}{\widehat}
\newcommand{\wt}{\widetilde}
\newcommand{\ts}{\textstyle}
\newcommand{\ds}{\displaystyle}
\newcommand{\scs}{\scriptstyle}
\newcommand{\vep}{\varepsilon}
\newcommand{\rhp}{\rightharpoonup}
\newcommand{\cl}{\circ\!\!\!\!\!-}
\newcommand{\bcs}{\dot{\,}.\dot{\,}}
\newcommand{\eqv}{\Leftrightarrow}
\newcommand{\leqv}{\Longleftrightarrow}
\newtheorem{co}{Corollary} 
\newtheorem{lm}{Lemma} 
\newtheorem{Ex}{Example} 
\newtheorem{Th}{Theorem}
\newtheorem{df}{Definition} 
\newtheorem{pr}{Property} 
\newtheorem{pro}{Proposition} 
\newtheorem{rem}{Remark} 

\newcommand{\auxan}{{X}^n}
\newcommand{\auxbn}{{U}^n}
\newcommand{\rvxn}{\empty}
\newcommand{\vauxa}{{\vc X}}
\newcommand{\vauxb}{{\vc U}}
\newcommand{\vx}{{\empty}}
\newcommand{\lvauxa}{{\lvc X}}
\newcommand{\lvauxb}{{\lvc U}}
\newcommand{\lvx}{{\empty}}

\newcommand{\lcv}{convex } 

\newcommand{\hugel}{{\arraycolsep 0mm
                    \left\{\ba{l}{\,}\\{\,}\ea\right.\!\!}}
\newcommand{\Hugel}{{\arraycolsep 0mm
                    \left\{\ba{l}{\,}\\{\,}\\{\,}\ea\right.\!\!}}
\newcommand{\HUgel}{{\arraycolsep 0mm
                    \left\{\ba{l}{\,}\\{\,}\\{\,}\vspace{-1mm}
                    \\{\,}\ea\right.\!\!}}
\newcommand{\huger}{{\arraycolsep 0mm
                    \left.\ba{l}{\,}\\{\,}\ea\!\!\right\}}}

\newcommand{\Huger}{{\arraycolsep 0mm
                    \left.\ba{l}{\,}\\{\,}\\{\,}\ea\!\!\right\}}}

\newcommand{\HUger}{{\arraycolsep 0mm
                    \left.\ba{l}{\,}\\{\,}\\{\,}\vspace{-1mm}
                    \\{\,}\ea\!\!\right\}}}

\newcommand{\hugebl}{{\arraycolsep 0mm
                    \left[\ba{l}{\,}\\{\,}\ea\right.\!\!}}
\newcommand{\Hugebl}{{\arraycolsep 0mm
                    \left[\ba{l}{\,}\\{\,}\\{\,}\ea\right.\!\!}}
\newcommand{\HUgebl}{{\arraycolsep 0mm
                    \left[\ba{l}{\,}\\{\,}\\{\,}\vspace{-1mm}
                    \\{\,}\ea\right.\!\!}}
\newcommand{\hugebr}{{\arraycolsep 0mm
                    \left.\ba{l}{\,}\\{\,}\ea\!\!\right]}}
\newcommand{\Hugebr}{{\arraycolsep 0mm
                    \left.\ba{l}{\,}\\{\,}\\{\,}\ea\!\!\right]}}

\newcommand{\HugebrB}{{\arraycolsep 0mm
                    \left.\ba{l}{\,}\\{\,}\vspace*{-1mm}\\{\,}\ea\!\!\right]}}

\newcommand{\HUgebr}{{\arraycolsep 0mm
                    \left.\ba{l}{\,}\\{\,}\\{\,}\vspace{-1mm}
                    \\{\,}\ea\!\!\right]}}

\newcommand{\hugecl}{{\arraycolsep 0mm
                    \left(\ba{l}{\,}\\{\,}\ea\right.\!\!}}
\newcommand{\Hugecl}{{\arraycolsep 0mm
                    \left(\ba{l}{\,}\\{\,}\\{\,}\ea\right.\!\!}}
\newcommand{\hugecr}{{\arraycolsep 0mm
                    \left.\ba{l}{\,}\\{\,}\ea\!\!\right)}}
\newcommand{\Hugecr}{{\arraycolsep 0mm
                    \left.\ba{l}{\,}\\{\,}\\{\,}\ea\!\!\right)}}

\newcommand{\hugepl}{{\arraycolsep 0mm
                    \left|\ba{l}{\,}\\{\,}\ea\right.\!\!}}
\newcommand{\Hugepl}{{\arraycolsep 0mm
                    \left|\ba{l}{\,}\\{\,}\\{\,}\ea\right.\!\!}}
\newcommand{\hugepr}{{\arraycolsep 0mm
                    \left.\ba{l}{\,}\\{\,}\ea\!\!\right|}}
\newcommand{\Hugepr}{{\arraycolsep 0mm
                    \left.\ba{l}{\,}\\{\,}\\{\,}\ea\!\!\right|}}

\newcommand{\MEq}[1]{\stackrel{
{\rm (#1)}}{=}}

\newcommand{\MLeq}[1]{\stackrel{
{\rm (#1)}}{\leq}}

\newcommand{\ML}[1]{\stackrel{
{\rm (#1)}}{<}}

\newcommand{\MGeq}[1]{\stackrel{
{\rm (#1)}}{\geq}}

\newcommand{\MG}[1]{\stackrel{
{\rm (#1)}}{>}}

\newcommand{\MPreq}[1]{\stackrel{
{\rm (#1)}}{\preceq}}

\newcommand{\MSueq}[1]{\stackrel{
{\rm (#1)}}{\succeq}}

\newenvironment{jenumerate}
	{\begin{enumerate}\itemsep=-0.25em \parindent=1zw}{\end{enumerate}}
\newenvironment{jdescription}
	{\begin{description}\itemsep=-0.25em \parindent=1zw}{\end{description}}
\newenvironment{jitemize}
	{\begin{itemize}\itemsep=-0.25em \parindent=1zw}{\end{itemize}}
\renewcommand{\labelitemii}{$\cdot$}

\newcommand{\iro}[2]{{\color[named]{#1}#2\normalcolor}}
\newcommand{\irr}[1]{{\color[named]{Red}#1\normalcolor}}
\newcommand{\irg}[1]{{\color[named]{Green}#1\normalcolor}}
\newcommand{\irb}[1]{{\color[named]{Blue}#1\normalcolor}}
\newcommand{\irBl}[1]{{\color[named]{Black}#1\normalcolor}}
\newcommand{\irWh}[1]{{\color[named]{White}#1\normalcolor}}

\newcommand{\irY}[1]{{\color[named]{Yellow}#1\normalcolor}}
\newcommand{\irO}[1]{{\color[named]{Orange}#1\normalcolor}}
\newcommand{\irBr}[1]{{\color[named]{Purple}#1\normalcolor}}
\newcommand{\IrBr}[1]{{\color[named]{Purple}#1\normalcolor}}
\newcommand{\irBw}[1]{{\color[named]{Brown}#1\normalcolor}}
\newcommand{\irPk}[1]{{\color[named]{Magenta}#1\normalcolor}}
\newcommand{\irCb}[1]{{\color[named]{CadetBlue}#1\normalcolor}}

%
\newenvironment{indention}[1]{\par
\addtolength{\leftskip}{#1}\begingroup}{\endgroup\par}
%
\newcommand{\namelistlabel}[1]{\mbox{#1}\hfill} 
\newenvironment{namelist}[1]{%
\begin{list}{}
{\let\makelabel\namelistlabel
\settowidth{\labelwidth}{#1}
\setlength{\leftmargin}{1.1\labelwidth}}
}{%
\end{list}}
%
%
\newcommand{\bfig}{\begin{figure}[t]}
\newcommand{\efig}{\end{figure}}
\setcounter{page}{1}

\newtheorem{theorem}{Theorem}

\newcommand{\ep}{\mbox{\rm e}}

\newcommand{\Exp}{\exp
}
\newcommand{\idenc}{{\varphi}_n}
\newcommand{\resenc}{
{\varphi}_n}
\newcommand{\ID}{\mbox{\scriptsize ID}}
\newcommand{\TR}{\mbox{\scriptsize TR}}
\newcommand{\Av}{\mbox{\sf E}}

\newcommand{\Vl}{|}
\newcommand{\Ag}{(R,P_{X^n}|W^n)}
\newcommand{\Agv}[1]{({#1},P_{X^n}|W^n)}
\newcommand{\Avw}[1]{({#1}|W^n)}

\newcommand{\Jd}{X^nY^n}
\newcommand{\IdR}{r_n}

\newcommand{\Index}{{n,i}}

\newcommand{\cid}{C_{\mbox{\scriptsize ID}}}
\newcommand{\cida}{C_{\mbox{{\scriptsize ID,a}}}}

\newcommand{\iN}{\rm (in)}
\newcommand{\ouT}{\rm (out)}

\arraycolsep 0.5mm
\date{}
%
\title{
Strong Converse Theorems for Degraded Broadcast Channels 
with Feedback
}
\author{%
\IEEEauthorblockA{
Yasutada Oohama\\
  University of Electro-Communications, Tokyo, Japan \\
  Email: oohama@uec.ac.jp} 
} 

\maketitle


\begin{abstract} 
We consider the discrete memoryless degraded broadcast 
channels with feedback. We prove that the error probability 
of decoding tends to one 
exponentially for rates outside the capacity region and 
derive an explicit lower bound of this exponent function. 
We shall demonstrate that the information spectrum approach 
is quite useful for investigating this problem.
\end{abstract}
%


\section{\normalsize DBC with Feedback}

\begin{figure}[t]
\bc
\includegraphics[width=7.2cm]{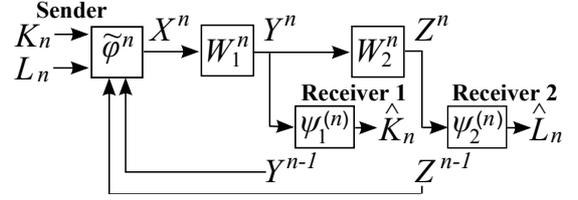}
\caption{Transmission of the message pair $(K_n,L_n)$ via the DBC with feedback.}
\label{fig:theGBCFb2} 
\ec
\end{figure}

\begin{figure}[t]
\bc
\includegraphics[width=7.2cm]{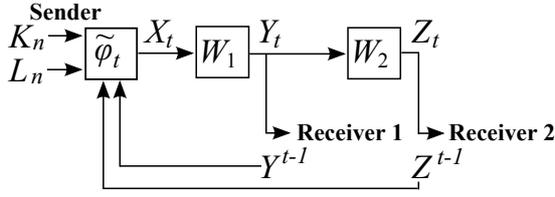}
\caption{The $t$-th transmission in the DBC with feedback.}
\label{fig:theGBCFb} 
\ec
\end{figure}

\begin{figure}[t]
\bc
\includegraphics[width=7.0cm]{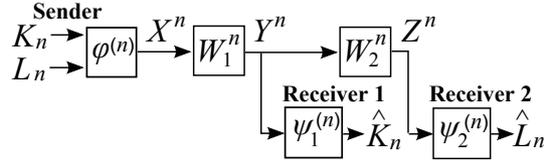}
\caption{Transmission of messages via the degraded BC.}
\label{fig:theGBC} 
\ec
\end{figure}

Let ${\cal X}, {\cal Y},$ and ${\cal Z}$ be finite sets.
The broadcast channel we study in this paper is defined 
by a discrete memoryless channel specified with the following 
stochastic matrix:
\beq
{W} \defeq \{{W}(y,z|x)\}_{
(x,y,z) 
\in    {\cal X}
\times {\cal Y} 
\times {\cal Z}}.
\eeq
Here ${\cal X}$ is a set of channel input and 
${\cal Y}$, and ${\cal Z}$ are sets of two channel outputs.
We assume that those are finite sets. Let $X^n$ be a random 
variable taking values in ${\cal X}^n$. 
We write an element of ${\cal X}^n$ as   
$x^n=x_{1}x_{2}$$\cdots x_{n}.$ 
Suppose that $X^n$ has a probability distribution on ${\cal X}^n$ 
denoted by 
$p_{X^n}=$ 
$\left\{p_{X^n}(x^n) \right\}_{{x^n} \in {\cal X}^n}$.
Similar notations are adopted for other random variables. 
Let $Y^n \in {\cal Y}^n$ and $Z^n \in {\cal Y}^n$  be random variables 
obtained as the channel output by connecting 
$X^n$ to the input of channel. 
We write a conditional distribution of $(Y^n,Z^n)$ on given $X^n$ 
as 
$$
W^n=
\left\{W^n(y^n,z^n|x^n)\right
\}_{(x^n,y^n,z^n)\in {\cal X}^n \times {\cal Y}^n \times {\cal Z}^n}.
$$
Since the channel is memoryless, we have 
\beq
W^n({y}^n,z^n|x^n)=\prod_{t=1}^nW (y_t,z_t|x_t).
\label{eqn:sde0}
\eeq
In this paper we deal with the case where the components 
$W({z},{y}|{x})$ of $W$ satisfy 
the following conditions:
\beq
W({y},{z}|{x})=W_1(y|x)W_2(z|y).
\label{eqn:sde1}
\eeq
In this case we say that the broadcast channel ${W}$ 
is {\it degraded}. The degraded broadcast channel (DBC) 
is specified by $(W_1,W_2)$. 
%
%
Let $K_n$ and  $L_n$ be uniformly 
distributed random 
variables taking values in message sets ${\cal K}_n $ and ${\cal L}_n$, 
respectively. 
The random variable $K_n$ is a message sent to the receiver 1.
The random variable $L_n$ is a message sent to the receiver 2.
In this paper we consider the case where we have feedback 
links from the receivers 1 and 2 to the sender. 
Transmission of the message pair $(K_n,L_n)$ 
via the DBC with feedback is shown in Fig. \ref{fig:theGBCFb2}.
A feedback encoder denoted by 
$\tilde{\varphi}^{n}$ 
$=\{\tilde{\varphi}_t\}_{t=1}^n$
consists of $n$ encoder functions 
$\tilde{\varphi}_t$, $t=1,2,\cdots,L$, 
where for each $t=1,2,$ $\cdots,n$, 
$$
\tilde{\varphi}_t: 
{\cal K}_n \times {\cal L}_n \times 
{\cal Y}^{t-1} \times {\cal Z}^{t-1} 
\to {\cal X}_t
$$
is a stochastic matrix. For a given message pair 
$(k,l)$ 
$\in {\cal K}_n \times {\cal L}_n$
and given feedback signals $y^{n-1}$ $\in {\cal Y}^n$ 
form the receiver 1  
and $z^{n-1}$ $\in {\cal Z}^n$ from the receiver 2, 
conditional provability of $x^n \in {\cal X}^n$ 
by $\tilde{\varphi}^n$ is  
$$
\tilde{\varphi}^n(x^n|k,l,y^{n-1},z^{n-1})
=
\prod_{t=1}^n \tilde{\varphi}_t(x_t|k,l,y^{t-1},z^{t-1}).
$$
The $t$-th transmission in the DBC with feedback
is shown in Fig. \ref{fig:theGBCFb}. 
The joint probability 
mass function on 
${\cal K}_n \times {\cal L}_n$ 
$\times {\cal X}^n$ 
$\times {\cal Y}^n$ 
$\times {\cal Z}^n$ 
is given by
\beqno
& &\Pr\{(K_n,L_n,X^n,Y^n,Z^n)=(k,l,x^n,y^n, z^n)\}
\nonumber\\
&=&\frac{1}{\pa{\cal K}_n\pa \pa{\cal L}_n\pa}
\prod_{t=1}^n\{
\tilde{\varphi}_t(x_t|k,l,y^{t-1},z^{t-1})
\nonumber\\
& &\qquad \qquad \qquad \times W_1\left(y_t\left|x_t\right.\right)
W_2\left(z_t\left|y_t\right.\right)\},
\eeqno
where $\pa {\cal K}_n \pa$ is a cardinality 
of the set ${\cal K}_n$. We set 
\beqno
& & \tilde{p}_{K_nL_nX^nY^nZ^n}(k,l,x^n,y^n,z^n)
\\
&\defeq&\Pr\{(K_n,L_n,X^n,Y^n,Z^n)=(k,l,x^n,y^n, z^n)\}.
\eeqno
By an elementary calculation we can show that
for each $(l,x^n,$ $y^n,z^n)$
$\in {\cal L}_n$ 
$\times {\cal X}^n$
$\times {\cal Y}^n$
$\times {\cal Z}^n$, the probability 
$
\tilde{p}_{L^nX^nY^nZ^n}($ $l,x^n,$ $y^n,z^n)
$
is given by 
\beqno
& &\tilde{p}_{L^nX^nY^nZ^n}(l,x^n,y^n,z^n)=\tilde{p}_{L^n}(l)
\\
&&\times \prod_{t=1}^n \left\{
\tilde{p}_{X_t|L_n X^{t-1}Y^{t-1}Z^{t-1}}
(x_t|l,x^{t-1},y^{t-1},z^{t-1}) \right.
\\
& &\qquad \times 
   W_1(y_t|x_t)W_2(z_t|y_t)\}.
\eeqno
The decoding functions 
at the receiver 1 and the receiver 2, respectively, 
are denoted by ${\psi}_1^{(n)}$ and ${\psi}_2^{(n)}$. 
Those functions are formally defined by
$
{\psi}_1^{(n)}: {\cal Y}^{n} \to {\cal K}_n,
{\psi}_2^{(n)}: {\cal Z}^{n} \to {\cal L}_n.
$
The average error probability of decoding on the receivers 1 and 2 
is defined by 
\beqno
& &{\rm P}_{\rm e, FB}^{(n)}=
   {\rm P}_{\rm e, FB}^{(n)}(\varphi^{(n)},\psi_1^{(n)},\psi_2^{(n)})
\\
& \defeq & \Pr\{\psi_{1}^{(n)}(Y^n)\neq K_n\mbox{ or } 
\psi_{2}^{(n)}(Z^n)\neq L_n \}
\eeqno
For $k\in {\cal K}_n$ and $l\in {\cal L}_n$, set
${\cal D}_1(k)\defeq $ $\{ y^n: \psi_1^{(n)}(y^n)=k \},$
${\cal D}_2(l)\defeq $ $\{ z^n: \psi_2^{(n)}(z^n)=l \}.$
The families of sets 
$\{ {\cal D}_1(k) \}_{k\in {\cal K}_n}$ and
$\{ {\cal D}_2(l) \}_{l \in {\cal L}_n}$ 
are called the decoding regions. 
Using the decoding region, 
${\rm P}_{\rm e, FB}^{(n)}$ can be written as
\beqno
& &{\rm P}_{\rm e, FB}^{(n)}
={\rm P}_{\rm e,FB}^{(n)}(\varphi^n,\psi_1^{(n)},\psi_2^{(n)})
\\
&=&\frac{1}{|{\cal K}_n| |{\cal L}_n|} 
\sum_{(k,l)\in {\cal K}_n \times {\cal L}_n }
\sum_{\scs (x^n,y^n,z^n)\in {\cal X}^n \times {\cal Y}^n\times {\cal Z}^n:
       \atop{
       \scs y^n \in {\cal D}_1(k)\mbox{ or }  
             \scs z^n  \in  {\cal D}_2(l)
       }
    }  
\\
& &\times 
\tilde{\varphi}^{n}(x^n|k,l,y^{n-1},z^{n-1})W_1^n(y^n|x^n)W_2^n(z^n|y^n).
\eeqno
The average correct probability of decoding 
is defined by 
$$
{\rm P}_{\rm c,FB}^{(n)}
={\rm P}_{\rm c, FB}^{(n)}(
\tilde{\varphi}^n,
\psi_1^{(n)},
\psi_2^{(n)})
=1-{\rm P}_{\rm e, FB}^{(n)}(\tilde{\varphi}^n,\psi_1^{(n)},\psi_2^{(n)}).
$$
\newcommand{\Zasss}{
This quantity has the following form:
\beqno
& &{\rm P}_{\rm c,FB}^{(n)}
={\rm P}_{\rm c, FB}^{(n)}(\tilde{\varphi}^n,\psi_1^{(n)},\psi_2^{(n)})
\\
&=&\frac{1}{|{\cal K}_n| |{\cal L}_n|} 
\sum_{(k,l)\in {\cal K}_n \times {\cal L}_n }
\sum_{\scs (x^n,y^n,z^n)\in {\cal X}^n \times {\cal Y}^n\times {\cal Z}^n:
       \atop{
       \scs y^n \in {\cal D}_1(k), 
             \scs z^n  \in  {\cal D}_2(l)
       }
    }  
\\
& &\times 
\tilde{\varphi}^{n}(x^n|k,l,y^{n-1}z^{n-1})
W_1^n(y^n|x^n)W_2^n(z^n|y^n).
\eeqno
}
On the other hand, transmission of messages 
via the DBC without feedback is shown 
in Fig. \ref{fig:theGBC}.
In this figure, $\varphi^{(n)}$ is 
a stochastic matrix given by
$$
\varphi^{(n)}=\{
\varphi^{(n)}(x^n|k,l)\}_{
(k,l,x^n)\in {\cal K}_n\times {\cal L}_n \times {\cal X}^n},
$$ 
where $\varphi^{(n)}(x^n|k,l)$ is a conditional probability 
of $x^n \in {\cal X}^n$ given message pair $(k,l)\in$
${\cal K}_n\times {\cal L}_n$.
Let the average error probability of decoding in the case without 
feedback be denoted by ${\rm P}^{(n)}_{\rm e}$.  
This quantity has the following form
\beqno
{\rm P}_{\rm e}^{(n)}
&=&\frac{1}{|{\cal K}_n| |{\cal L}_n|} 
\sum_{(k,l)\in {\cal K}_n \times {\cal L}_n }
\sum_{\scs (x^n,y^n,z^n)\in {\cal X}^n \times {\cal Y}^n\times {\cal Z}^n:
       \atop{
       \scs y^n   \in {\cal D}_1(k) \mbox{ or }  
           \scs z^n  \in  {\cal D}_2(l)
       }
    }  
\\
& &\times 
\varphi^{(n)}(x^n|k,l)W_1^n(y^n|x^n)W_2^n(z^n|y^n).
\eeqno
The average correct probability of decoding 
is defined by 
\beqno
{\rm P}^{(n)}_{\rm c}=
   {\rm P}^{(n)}_{\rm c}(\varphi^{(n)},\psi_1^{(n)},\psi_2^{(n)})
\defeq 
1-{\rm P}^{(n)}_{\rm e}(\varphi^{(n)}, \psi_1^{(n)},\psi_2^{(n)}).
\eeqno

For $\varepsilon$ $\in (0,1)$, a pair $(R_1,R_2)$ is 
$\varepsilon$-{\it achievable} if there exists a sequence 
of triples $\{(\tilde{\varphi}^{n},$ 
$\psi_1^{(n)}, \psi_2^{(n)})\}_{n=1}^{\infty}$ 
such that 
\beqa 
& &{\rm P}_{{\rm e,FB}}^{(n)}
(\tilde{\varphi}^{n},\psi_1^{(n)},\psi_2^{(n)})
\leq \varepsilon, 
\nonumber\\
& &\liminf_{n\to\infty} \nbn \log \pa {\cal K}_n \pa  \geq  R_1,
\liminf_{n\to\infty} \
\nbn \log \pa {\cal L}_n \pa \geq R_2.
\nonumber
\eeqa
The set that consists of all $\varepsilon$-achievable rate pair is denoted by 
${\cal C}_{\rm DBC,FB}(\varepsilon|W_1,W_2)$. Furthermore, set 
$$
{\cal C}_{\rm DBC,FB}(W_1,W_2)
=\bigcap_{\epsilon\in (0,1)}
{\cal C}_{\rm DBC,FB}(\varepsilon|W_1,W_2).
$$

We define the capacity region 
${\cal C}_{\rm DBC}(\varepsilon| W_1,W_2)$ 
in the case without feedback 
in a manner quite similar to the definition of 
${\cal C}_{\rm DBC,FB}(\varepsilon| W_1,W_2)$.
We define the capacity region ${\cal C}_{\rm DBC}(W_1,W_2)$
of the DBC without feedback in a manner 
quite similar to the definition of 
${\cal C}_{\rm DBC, FB}(W_1,W_2)$. 

To describe ${\cal C}_{\rm DBC}(W_1,W_2)$, 
we introduce an auxiliary random variable $U$ taking 
values in a finite set ${\cal U}$. We assume 
that the joint distribution of 
$(U,X,Y,Z)$ is 
\beqno 
& &p_{U{X}{Y}Z}(u,x,y,z)
\\
&=&p_{U}(u)p_{X|U}(x|u)
W_1(y|x)W_2(z|y). 
\eeqno
The above condition is equivalent to $U \markov 
X \leftrightarrow Y $ $\markov Z$. Define the set of probability 
distribution $p=p_{UXYZ}$  of $(U,$ $X,$ $Y,$ $Z)$ $\in$ ${\cal U}$
$\times{\cal X}$ $\times{\cal Y}$ $\times{\cal Z}$
by
\beqno
&&{\cal P}(W_1,W_2)
\defeq 
\{p: \pa {\cal U} \pa 
\leq \pa {\cal X} \pa + 1,
\vSpa\\
& &\quad p_{Y|X}=W_1, p_{Z|Y}=W_2, 
U \markov  X\markov Y \markov Z \}.
\eeqno
Set 
\beqno
{\cal C}(p)
&\defeq &
\ba[t]{l}
\{(R_1,R_2): R_1,R_2 \geq 0\,,
\vSpa\\
\ba{rcl}
R_1 & \leq & I_p(X;Y|U), 
R_2  \leq I_p(U;Z)\}.
\ea
\ea
\\
{\cal C}(W_1,W_2)
&=& \bigcup_{p\in {\cal P}(W_1,W_2)}
{\cal C}(p).
\eeqno
\newcommand{\Zap}{
We can show that the above functions and sets 
satisfy the following property. 
\begin{pr}\label{pr:pro0}  
$\quad$
\begin{itemize}
\item[a)] The region ${\cal C}(W_1,W_2)$ is a closed convex set. 

\item[b)] The region ${\cal C}(W_1,W_2)$ can be expressed 
with a family of supporting hyperplanes. To describe this result 
we define the set of probability distribution 
$p=p_{UXYZ}$ of $(U,$ $X,$ $Y,$ $Z)$ $\in$ ${\cal U}$
$\times{\cal X}$ $\times{\cal Y}$ $\times{\cal Z}$ by
\beqno
&&{\cal P}_{\rm sh}(W_1,W_2)
\defeq 
\{p: 
\pa {\cal U} \pa \leq \pa {\cal X} \pa, 
\vSpa\\
&&\quad p_{Y|X}=W_1, p_{Z|Y}=W_2, 
U \markov  X\markov Y \markov Z \}.
\eeqno
We set
\beqno
& &C^{(\mu)}(W_1,W_2) 
\\
&\defeq& 
\max_{p \in {\cal P}_{\rm sh}(W_1,W_2) }
\left\{\mu I_p(X;Y|U)+I_p(U;Z)\right\},
\\
&&{\cal C}(W_1,W_2)
\\
&=& 
\bigcap_{\mu>0}\{(R_1,R_2):\mu R_1+R_2 
\leq C^{(\mu)}(W_1,W_2)\}.
\eeqno 
Then we have the following
\beqno
{\cal C}(W_1,W_2)={\cal C}_{\rm sh}(W_1,W_2).
\eeqno
\end{itemize}
\end{pr}
}

The broadcast channel was posed investigated 
by Cover $\cite{cov72}$.
Previous results on the capacity region for the DBC are 
given by the following theorem. 
\begin{Th}[\cite{bgm73}-
\cite{agk76}]\label{th:ddirect}
For each fixed $\varepsilon \in (0,1)$ and any DBC 
$(W_1,W_2)$, we have 
\beqno
&&{\cal C}_{\rm DBC}(\varepsilon|W_1,W_2)
 ={\cal C}_{\rm DBC}(W_1,W_2)
\\
&=&{\cal C}(W_1,W_2).
\eeqno
\end{Th}

A previous result on ${\cal C}_{\rm DBC,FB}($$W_1,W_2)$ 
is given by the following theorem stating that the feedback can not 
increase the capacity region for the DBC.
\begin{Th}[El Gamal \cite{ElGamalBcFb78}]\label{th:fbcapreg}
{For any DBC $(W_1,W_2$ $)$, we have
\beqno
{\cal C}_{\rm DBC,FB}(W_1,W_2)
&=&{\cal C}_{\rm DBC}(W_1,W_2)
\\
&=&{\cal C}(W_1,W_2).
\eeqno
}
\end{Th}

In general broadcast channels the feedback can increase 
the capacity region. Previous works on the coding problem for 
broad cast channels with feedback are summarized in 
\cite{ShaWi13BcFb13}. 

To examine an asymptotic behavior of ${\rm P}_{\rm c,FB}^{(n)}$ 
for rate pairs outside the capacity region 
we define the following quantity. 
\beqno
& & 
G^{(n)}_{\rm FB}(R_1,R_2|W_1,W_2)
\\
&\defeq&
\min_{\scs 
(\tilde{\varphi}^{n},\psi_1^{(n)},\psi_2^{(n)}):
    \atop{\scs 
         (1/n)\log | {\cal K}_n |\geq R_1,
         \atop{\scs 
         (1/n)\log | {\cal L}_n |\geq R_2
         }
    }
}
\hspace*{-4mm}
\left(-\frac{1}{n}\right)
\log {\rm P}_{\rm c,FB}^{(n)}
(\tilde{\varphi}^{n},\psi_1^{(n)},\psi_2^{(n)}),
\\
& &
G_{\rm FB}(R_1,R_2|W_1,W_2)
=\lim_{n \to \infty}G^{(n)}_{\rm FB}(R_1,R_2|W_1,W_2).
\eeqno
The quantity $G_{\rm FB}(R_1,R_2|W_1,W_2)$
is the optimal exponent function for 
the correct probability of decoding at rate pairs 
outside the capacity region. In the case without 
feedback we define the optimal exponent function 
$G_{\rm }(R_1,R_2|W_1,W_2)$ 
for the correct probability of decoding for rate pairs 
outside the capacity region in a manner quite 
similar to the definition of $G_{\rm FB}(R_1,R_2|W_1,W_2)$. 

Define 
\beqno
& &\omega^{(\mu)}_{q}(x,y,z|u)
\\
&\defeq&\mu\log \frac{q_{Y|X}(y|x) }{q_{Y|U}(y|u)}
+\log \frac{q_{Z|U}(z|u)}{q_{Z}(z)},
\\
& &\Lambda_{q}^{(\mu,\lambda)}({XYZ|U})
\\
&\defeq &\sum_{(u,x,y,z)\in 
{\cal U} \times 
{\cal X} \times 
{\cal Y} \times 
{\cal Z}
}q_{UX}(u,x)q_{Y|X}(y|x)q_{Z|Y}(z|y)
\\
&&\quad \times
\exp\left\{\lambda \omega^{(\mu)}_{q}(x,y,z|u) \right\},
\\
& &\Omega^{(\mu,\lambda)}_{q}(XYZ|U)
\defeq\log \Lambda_{q}^{(\mu,\lambda)}({XYZ|U}),
\\
& &\Omega^{(\mu,\lambda)}(W_1,W_2)
\defeq \max_{q \in{\cal P}(W_1,W_2)}
\Omega_{q}^{(\mu,\lambda)}({XYZ|U}),
\\
& &
F^{(\mu,\lambda)}(\mu R_1+R_2|W_1,W_2)
\\
&\defeq&
\frac{\lambda(\mu R_1+R_2)-\Omega^{(\mu,\lambda)}(W_1,W_2)}
{1+2\lambda+\lambda\mu},
\\
\\
& &F(R_1,R_2|W_1,W_2)
\\
&\defeq&\sup_{\mu,\lambda >0} 
\frac{\lambda(\mu R_1+R_2)
-\Omega^{(\mu,\lambda)}(W_1,W_2)}
{1+2\lambda+\lambda\mu}.
\eeqno

We can show that the above functions and sets 
satisfy the following property. 
\begin{pr}\label{pr:pro1}  
$\quad$
\begin{itemize}
\item[a)]
For each $q \in {\cal P}(W_1,W_2)$, 
$\Omega_{q}^{(\mu,\lambda)}(XYZ|U)$ is a monotone increasing and 
convex function of $\lambda>0$. 
\item[b)] For every $q \in {\cal P}(W_1,W_2)$, we have 
\beqno
\lim_{\lambda\to +0}
\frac{\Omega_{q}^{(\mu,\lambda)}(XYZ|U)}{\lambda}
&=&\mu I_q(X;Y|U)+I_q(U;Z).
\eeqno
\item[c)] If $(R_1,R_2) \notin {\cal C}(W_1,W_2)$,
then we have $F(R_1,$ $R_2|$ $W_1,W_2)>0$. 
\end{itemize}
\end{pr}

The author \cite{OohamaDbcISIT15} obtained the following. 
\begin{Th}\label{Th:main} 
For any DBC $(W_1,W_2)$, we have 
\beqa
G(R_1,R_2|W_1,W_2) &\geq& F(R_1,R_2|W_1,W_2).
\label{eqn:mainIeq}
\eeqa
\end{Th}

It follows from Theorem \ref{Th:main} and Property 
\ref{pr:pro1} part c) that if $(R_1,R_2)$ is outside 
the capacity region, then the error probability of 
decoding goes to one exponentially and its exponent 
is not below $F(R_1,R_2|W_1,W_2)$.
%
%

Our result in the case of feedback is the following. 
\begin{Th}\label{Th:Fbmain}
For any DBC $(W_1,W_2)$, we have 
\beqa
G_{\rm FB}(R_1,R_2|W_1,W_2) &\geq& {F}(R_1,R_2|W_1,W_2).
\label{eqn:FbmainIeq}
\eeqa
\end{Th}

It is interesting that the exponent function $F(R_1,R_2$ $|W_1,W_2)$ also serves as 
a lower bound of the optimal exponent function   
$G_{\rm FB}(R_1,R_2|W_1,W_2)$ in the case of feedback. This result strongly 
suggests a possibility that the feedback can not improve the optimal 
exponent function for the probability of correct decoding 
at the rate pairs outside the capacity region. 

From this theorem we immediately follows from the following
corollary. 
\begin{co} For each fixed $\varepsilon$ $ \in (0,1)$,
and any DBC $(W_1,W_2)$, we have 
\beqno
&&{\cal C}_{\rm DBC,FB}(\varepsilon|W_1,W_2)
 ={\cal C}_{\rm DBC}(\varepsilon|W_1,W_2)
\\
&=&{\cal C}_{\rm DBC}(W_1,W_2)={\cal C}(W_1,W_2).
\eeqno
\end{co}

Outline of the proof of Theorem \ref{Th:Fbmain} will be given 
in the next section. The exponent function at rates outside 
the channel capacity in the case without feedback was derived by 
Arimoto \cite{ari} and Dueck and K\"orner \cite{dk}. 
The exponent function at rates outside 
the channel capacity in the case with feedback was derived by 
Csisz\'ar and K\"orner \cite{ckf}. They show that feedback 
can not improve the 
reliability function for the DMC at rates above capacity. 
The techniques used by them are not sufficient to prove 
Theorem \ref{Th:main}. Some novel techniques based on 
the information spectrum method introduced by Han \cite{han} 
are necessary to prove this theorem.

\section{Outline of the Proof of the Main Result}  

In this section we outline the proof of Theorem \ref{Th:Fbmain}. 
We first prove the following lemma. 
\begin{lm}\label{lm:FbOhzzz}
For any $\eta>0$ and for any 
$(\tilde{\varphi}^{n},\psi_1^{(n)},\psi_2^{(n)})$  
satisfying 
$
(1/n) \log |{\cal K}_n| \geq R_1,
$ $
(1/n) \log |{\cal L}_n| \geq R_2,
$
we have 
\beqa	
&   &{\rm P}_{\rm c,FB}^{(n)}
(\tilde{\varphi}^{n},\psi_1^{(n)},\psi_2^{(n)})
\leq \tilde{p}_{L_nX^nY^nZ^n}\hugel
\nonumber\\
& &
R_1\leq \frac{1}{n}\log
\frac{W_1^n(Y^n|X^n)W_2^n(Z^n|Y^n)}{q_{Y^nZ^n|L_n}(Y^n,Z^n|L_n)}+\eta,
\label{eqn:Fasppb}\\
& &
R_2 \leq \left.
\frac{1}{n}\log
\frac{\tilde{p}_{Z^n|L_n}(Z^n|L_n)}{\tilde{q}_{Z^n}(Z^n)}+\eta
\right\}
+2{\rm e}^{-n\eta}. 
\label{eqn:Fbazsad}
\eeqa
%
In (\ref{eqn:Fasppb}), we can choose 
any conditional distribution $q_{Y^nZ^n|L_n}$ on 
${\cal Y}^n\times{\cal Z}^n$
given $L_n$ $\in {\cal L}_n$. 
In (\ref{eqn:Fbazsad}) we can choose any probability 
distribution $\tilde{q}_{Z^n}$ on ${\cal Z}^n$. 
\end{lm}

Proof of this lemma is given in Appendix \ref{sub:Apda}.
\newcommand{\ApdaFb}{
\subsection{Proof of Lemma \ref{lm:FbOhzzz}}
\label{sub:Apda}

In this appendix we prove Lemma \ref{lm:FbOhzzz}. 

{\it Proof of Lemma \ref{lm:FbOhzzz}:} 
For $l\in {\cal L}_n$, set
\beqno
\tilde{\cal A}_1(l)& \defeq &
\{(x^n, y^n,z^n): 
\ba[t]{l}
W_2^n(z^n|y^n)W_1^n(y^n|x^n)
\\
\geq |{\cal K}_n| {\rm e}^{-n\eta}
q_{Y^nZ^n|L_n}(y^n,z^n|l)
\},
\ea
\\
\tilde{\cal A}_2(l)
&\defeq &
\{(x^n,y^n,z^n): 
\ba[t]{l}
\tilde{p}_{Z^n|L_n}(z^n|l)
\geq |{\cal L}_n| {\rm e}^{-n\eta}
\tilde{q}_{Z^n}(z^n)\},
\ea
\\
\tilde{\cal A}(l)&\defeq&
\tilde{\cal A}_1(l)\cap \tilde{\cal A}_2(l). 
\eeqno
Then we have the following: 
\beqno
{\rm P}_{\rm c,FB}^{(n)}
&=&\frac{1}{|{\cal K}_n| |{\cal L}_n|} 
\sum_{(k,l)\in {\cal K}_n \times {\cal L}_n }
\sum_{ {\scs (x^n,y^n,z^n)\in \tilde{\cal A}(l),
         \atop{
          \scs y^n   \in {\cal D}_1(k), 
          \scs z^n  \in  {\cal D}_2(l)
         }
     }
   }1  
\\
& &\times 
\tilde{p}_{X^nY^nZ^n|K_n,L_n}(x^n,y^n,z^n|k,l)
\\
&&+\frac{1}{|{\cal K}_n| |{\cal L}_n|} 
  \sum_{(k,l)\in {\cal K}_n \times {\cal L}_n }
  \sum_{\scs (x^n,y^n,z^n)\in \tilde{\cal A}^c(l):
       \atop{
       \scs y^n \in {\cal D}_1(k), 
            z^n \in {\cal D}_2(l)
       }
    }1 
\\
& &\times 
\tilde{p}_{X^nY^nZ^n|K_n,L_n}(x^n,y^n,z^n|k,l)
\\
&\leq& \sum_{i=0,1,2} \tilde{\Delta}_i,
\eeqno
where
\beqno
\tilde{\Delta}_0
&\defeq &\frac{1}{|{\cal K}_n| |{\cal L}_n|} 
\sum_{(k,l)\in {\cal K}_n \times {\cal L}_n }
\sum_{\scs (x^n,y^n,z^n)\in \tilde{\cal A}(l)}1
\\
& &\times 
\tilde{p}_{X^nY^nZ^n|K_n,L_n}(x^n,y^n,z^n|k,l),
\\
\tilde{\Delta}_i
&\defeq &
  \frac{1}{|{\cal K}_n| |{\cal L}_n|} 
  \sum_{(k,l)\in {\cal K}_n \times {\cal L}_n }
  \sum_{\scs (x^n,y^n,z^n)\in \tilde{\cal A}_i^c(l),
       \atop{\scs 
            y^n \in {\cal D}_1(k), 
            z^n \in {\cal D}_2(l)
       }
    }1
\\
& &\times 
\tilde{p}_{X^nY^nZ^n|K_n,L_n}(x^n,y^n,z^n|k,l)
\\
& &\mbox{ for }i=1,2.
\eeqno
By definition we have 
\beqa
& &\tilde{\Delta}_0
\nonumber\\
&=&\tilde{p}_{L_nX^nY^nZ^n}\hugel
\nonumber\\
& &
\frac{1}{n}\log |{\cal K}_n| \leq \frac{1}{n}\log
\frac{W_1^n(Y^n|X^n)W_2^n(Z^n|Y^n)}{q_{Y^nZ^n|L_n}(Y^n,Z^n|L_n)}+\eta,
\nonumber\\
& &
\frac{1}{n}\log |{\cal L}_n| \leq \left.
\frac{1}{n}\log\frac{\tilde{p}_{Z^n|L_n}(Z^n|L_n)}{\tilde{q}_{Z^n}(Z^n)}+\eta
\right\}.
\label{eqn:Fbazsadaba}
\eeqa
From (\ref{eqn:Fbazsadaba}), it follows that 
if $(\varphi^{(n)},\psi_1^{(n)},\psi_2^{(n)})$  
satisfies   
$$
\frac{1}{n}\log |{\cal K}_n| \geq R_1,
\frac{1}{n}\log |{\cal L}_n| \geq R_2,
$$
then the quantity $\tilde{\Delta}_0$ is upper bounded by 
the first term in the right members of (\ref{eqn:Fbazsad}) 
in Lemma \ref{lm:FbOhzzz}.
Hence it suffices to show 
$\tilde{\Delta}_i\leq {\rm e}^{-n\eta},i=1,2$ 
to prove Lemma \ref{lm:FbOhzzz}. 
We first prove $\tilde{\Delta}_1\leq {\rm e}^{-n\eta}$. 
We have the following chain of inequalities: 
\beqno
\tilde{\Delta}_1
&= &
  \frac{1}{|{\cal K}_n| |{\cal L}_n|} 
  \sum_{(k,l)\in {\cal K}_n \times {\cal L}_n }
  \sum_{\scs 
        (x^n,y^n,z^n):
         \atop{\scs 
             y^n \in {\cal D}_1(k), 
             z^n \in {\cal D}_2(l)
                \atop{\scs 
                    W_1^n(y^n|x^n)W_2(z^n|y^n)
                    \atop{\scs   
                     <{\rm e}^{-n\eta}|{\cal K}_n| 
                         \atop{\scs 
                         \times q_{Y^nZ^n|L_n}(y^n,z^n|l)
                         }
                     }
                }
         }
    }1
\\
& &\times 
\tilde{\varphi}^{n}(x^n|k,l,y^{n-1},z^{n-1})
W_1^n(y^n|x^n)W_2^n(z^n|y^n) 
\\
&\leq &
  \frac{ {\rm e}^{-n\eta} }{|{\cal L}_n|} 
  \sum_{(k,l)\in {\cal K}_n \times {\cal L}_n }
  \sum_{\scs
         (x^n,y^n,z^n):
           \atop {\scs 
            y^n \in {\cal D}_1(k), 
            z^n \in {\cal D}_2(l)
       }
    }1 
\\
& &\times 
\tilde{\varphi}^{n}(x^n|k,l,y^{n-1},z^{n-1})
q_{Y^nZ^n|L_n}(y^n,z^n|l) 
\\
&=&
  \frac{ {\rm e}^{-n\eta} }{|{\cal L}_n|} 
  \sum_{(k,l)\in {\cal K}_n \times {\cal L}_n }
q_{Y^nZ^n|L_n}\left(
\left.{\cal D}_1(k)\times {\cal D}_2(l)
\right| l \right)  
\\
&\leq&
  \frac{ {\rm e}^{-n\eta} }{|{\cal L}_n|} 
  \sum_{ l \in {\cal L}_n }
  \sum_{ k\in {\cal K}_n }
  q_{Y^n|L_n}\left(\left.{\cal D}_1(k)\right| l \right) 
\\
&=&
  \frac{ {\rm e}^{-n\eta} }{|{\cal L}_n|} 
  \sum_{ l\in {\cal L}_n }
  q_{Y^n|L_n}
\left(\left. \bigcup_{k\in {\cal K}_n}{\cal D}_1(k)\right|l \right) 
\\
&\leq&
  \frac{ {\rm e}^{-n\eta} }{|{\cal L}_n|} 
  \sum_{ l\in {\cal L}_n }1= {\rm e}^{-n\eta}.
\eeqno
Next we prove $\tilde{\Delta}_2\leq {\rm e}^{-n\eta}$.
We have the following chain of inequalities:
\beqno
\tilde{\Delta}_2
&= &
  \frac{1}{|{\cal L}_n|} 
  \sum_{(k,l) \in {\cal K}_n \times {\cal L}_n }
  \sum_{\scs 
        (x^n,y^n,z^n):
         \atop{\scs 
             y^n \in {\cal D}_1(k), 
             z^n \in {\cal D}_2(l)
                \atop{\scs 
                    \tilde{p}_{Z^n|L_n}<{\rm e}^{-n\eta}
                    \atop{\scs   
                      \times |{\cal L}_n| \tilde{q}_{Z^n}(z^n)
                    }
                }
         }
    }1
\\
& &\times 
\tilde{p}_{K_nX^nY^nZ^n|L_n}(k,x^n, y^n, z^n|l) 
\\
&\leq & 
  \frac{1}{|{\cal L}_n|} \sum_{l\in {\cal L}_n} 
  \sum_{\scs 
        z^n \in {\cal D}_2(l),
                \atop{\scs 
                    \tilde{p}_{Z^n|L_n}<{\rm e}^{-n\eta}
                    \atop{\scs   
                      \times |{\cal L}_n| \tilde{q}_{Z^n}(z^n)
                    }
                }
    }
\sum_{k\in {\cal K}_n} 
\sum_{(x^n,y^n)\in {\cal X}^n\times {\cal Y}^n }1
\\
& &\times 
\tilde{p}_{K_nX^nY^nZ^n|L_n}(k,x^n, y^n, z^n|l) 
\\
&\leq & 
  \frac{1}{|{\cal L}_n|} 
  \sum_{l\in {\cal L}_n} 
  \sum_{\scs 
            z^n \in {\cal D}_2(l),
            \atop{\scs 
                \tilde{p}_{Z^n|L_n}(z^n|l) <{\rm e}^{-n\eta}
                   \atop{\scs   
                        \times |{\cal L}_n| \tilde{q}_{Z^n}(z^n)
                    }
                }
        }  
    \tilde{p}_{Z^n|L_n}(z^n|l) 
\\
&\leq & 
  {\rm e}^{-n\eta} 
   \sum_{l\in {\cal L}_n }\sum_{z^n \in {\cal D}_2(l)}\tilde{q}_{Z^n}(z^n)
\\
&=& {\rm e}^{-n\eta} 
   \sum_{l\in {\cal L}_n }\tilde{q}_{Z^n}\left({\cal D}_2(l)\right)
\\
&=&{\rm e}^{-n\eta}\tilde{q}_{Z^n}
\left(\bigcup_{l \in {\cal L}_n} {\cal D}_2(l) \right)\leq {\rm e}^{-n\eta}. 
\eeqno
Thus Lemma \ref{lm:FbOhzzz} is proved 
\hfill\IEEEQED
}
%
For $t=1,2,$ $\cdots,n$, set 
\beqno
& & {\cal U}_t\defeq {\cal L}_n 
\times {\cal Y}^{t-1} 
\times {\cal Z}^{t-1}, 
{\cal V}_t\defeq {\cal L}_n \times {\cal Z}^{t-1},
\\
& &
{U}_t \defeq (L_n,Y^{t-1},Z^{t-1}) \in {\cal U}_t,
{V}_t \defeq (L_n,Z^{t-1}) \in {\cal V}_t,
\\
& & {u}_t \defeq (l,y^{t-1},z^{t-1})\in {\cal U}_t, 
    {v}_t \defeq (l,z^{t-1})\in {\cal V}_t. 
\eeqno
For each $t=1,2\cdots,l$, let $\kappa_t$ 
be a natural projection from 
${\cal U}_t$ onto ${\cal V}_t$.  
Using $\kappa_t$, we have $V_t=
\kappa_t(U_t),$ $t=1,2,\cdots,n$.
For each $t=1,2,\cdots,n$, let 
$
\tilde{\cal Q}(\empty{\cal U}_t $ 
$\times {\cal X} \times $
${\cal Y} \times {\cal Z})
$
be a set of all probability distributions on
$$ 
\empty{\cal U}_t \times {\cal X} \times 
   {\cal Y} \times {\cal Z}
={\cal L}_n \times {\cal X} \times {\cal Y}^t \times {\cal Z}^t.
$$
For $t=1,2,\cdots, n$, we simply write  
$\tilde{\cal Q}_t$$=$$\tilde{\cal Q}(\empty{\cal U}_t $ 
$\times {\cal X} \times {\cal Y} \times {\cal Z})$.
Similarly, for $t=1,2,\cdots, n$, we simply write 
$\tilde{q}_t=$ $\tilde{q}_{U_tX_tY_tZ_t}$ 
$\in \tilde{\cal Q}_t$.
Set
\beqno
\tilde{\cal Q}^n&\defeq& 
\prod_{t=1}^n \tilde{\cal Q}_t=\prod_{t=1}^n \tilde{\cal Q}(
\empty{\cal U}_t\times {\cal X}\times{\cal Y}\times {\cal Z}),
\\
\tilde{q}^n & \defeq & \left\{ \tilde{q}_t \right\}_{t=1}^n 
\in \tilde{\cal Q}^n.
\eeqno

\newcommand{\ApdLemmaA}{
}{

From Lemma \ref{lm:FbOhzzz}, we have the following lemma 
\begin{lm}\label{lm:FbOhzzzB}
For any $\eta>0$ and for any 
$(\varphi^{(n)},\psi_1^{(n)},\psi_2^{(n)})$  
satisfying  
$$
\frac{1}{n}\log |{\cal K}_n| \geq R_1,
\frac{1}{n}\log |{\cal L}_n| \geq R_2.
$$
we have
\beqa
& &{\rm P}_{\rm c,FB}^{(n)}
(\varphi^{(n)},\psi_1^{(n)},\psi_2^{(n)})
 \leq \tilde{p}_{L_n X^nY^nZ^n}
\hugel
\nonumber\\
& &{
R_1\leq \frac{1}{n}
\sum_{t=1}^n
\log \frac{W_1(Y_t|X_t)}{q_{Y_t|L_nY^{t-1}}(Y_t|L_n,Y^{t-1},Z^{t-1})}+\eta,}
\nonumber\\
& &
R_2\leq
\frac{1}{n}
\sum_{t=1}^n\log
\frac{\tilde{p}_{Z_t|{L_n}Z^{t-1}}
(Z_t|L_n,Z^{t-1})}{\tilde{q}_{Z_t}(Z_t)}
+\eta 
\huger +2{\rm e}^{-n\eta}.
\nonumber
\eeqa
\end{lm}

{\it Proof:} 
In (\ref{eqn:Fasppb}) in Lemma \ref{lm:FbOhzzz}, 
we choose $q_{Z^nY^n|L_n}$ 
\beqno
& &q_{Y^nZ^n|L_n}(y^n,z^n|l)
\\
&=&\prod_{t=1}^n 
\left\{
q_{Y_t|L_n Y^{t-1} Z^{t-1}}(y_t|l,y^{t-1},z^{t-1}) \right.
\\ 
& & \qquad \times \left. q_{ Z_t| L_n Y^t Z^{t-1} }(z_t|l,y^t,z^{t-1}) \right\}
\\
&=&\prod_{t=1}^n \{q_{Y_t|L_n Y^{t-1} Z^{t-1}}
(y_t|l,y^{t-1},z^{t-1})W_2(z_t|y_t)\}.
\eeqno
In (\ref{eqn:Fbazsad}) in Lemma \ref{lm:FbOhzzz}, 
we choose $\tilde{q}_{Z^n}$ having the form 
$$
\tilde{q}_{Z^n}(Z^n)=\prod_{t=1}^n\tilde{q}_{Z_t}(Z_t).
$$
Then from the bound (\ref{eqn:Fbazsad}) 
in Lemma \ref{lm:FbOhzzz}, we obtain 
\beqa
& &{\rm P}_{\rm c,FB}^{(n)}
(\varphi^{(n)},\psi_1^{(n)},\psi_2^{(n)})
\leq \tilde{p}_{L_nX^nY^nZ^n}\hugel
\nonumber\\
& &
R_1\leq \frac{1}{n}
\sum_{t=1}^n
\log \frac{W_1(Y_t|X_t)}{q_{Y_t|L_nY^{t-1}Z^{t-1}}
(Y_t|L_n,Y^{t-1},Z^{t-1})}+\eta,
\nonumber\\
& &
R_2\leq
\frac{1}{n}
\sum_{t=1}^n\log
\frac{\tilde{p}_{Z_t|{L_n}Z^{t-1}}(Z_t|L_n,Z^{t-1})}
{\tilde{q}_{Z_t}(Z_t)}
+\eta \huger +2{\rm e}^{-n\eta},
\nonumber
\eeqa
completing the proof. \hfill\IEEEQED

From Lemma \ref{lm:FbOhzzzB}, we immediately obtain 
the following lemma.
}

\begin{lm}\label{lm:FbOhzzzb}
For any $\eta>0$, for any 
$(\tilde{\varphi}^{n},\psi_1^{(n)},\psi_2^{(n)})$  
satisfying 
$$
\frac{1}{n}\log |{\cal K}_n| \geq R_1,
\frac{1}{n}\log |{\cal L}_n| \geq R_2,
$$
and for any $\tilde{q}^n \in \tilde{\cal Q}^n$, 
we have
\beqa
& &{\rm P}_{\rm c,FB}^{(n)}
(\tilde{\varphi}^{n},\psi_1^{(n)},\psi_2^{(n)})
\leq \tilde{p}_{L_nX^nY^nZ^n}
\hugel
\nonumber\\
& &{
R_1\leq \frac{1}{n}
\sum_{t=1}^n
\log \frac{W_1(Y_t|X_t)}{\tilde{q}_{Y_t|\empty{U}_t}(Y_t|\empty{U}_t)}
+\eta,}
\nonumber\\
& &
R_2\leq
\frac{1}{n}
\sum_{t=1}^n\log
\frac{\tilde{p}_{Z_t|
\empty{V}_t}(Z_t|\empty{V}_t)}{\tilde{q}_{Z_t}(Z_t)}
+\eta \huger +2{\rm e}^{-n\eta},
\label{eqn:saQQa}
\eeqa
where for each $t=1,2,\cdots,n$, the conditional probability 
distribution $\tilde{q}_{Y_t|U_t}$ and the probability 
distribution $\tilde{q}_{Z_t}$ appearing in the first term 
in the right members of (\ref{eqn:saQQa}) are chosen so that 
they are induced by the joint distribution 
$\tilde{q}_t=\tilde{q}_{U_tX_tY_tZ_t}$
$\in \tilde{\cal Q}_t$. 
\end{lm}

%
%
Here we define a quantity which serves as an exponential
upper bound of (\ref{eqn:saQQa}) in Lemma \ref{lm:FbOhzzzb}.
To describe this quantity we define some sets 
of probability distributions. 
Let ${\cal P}_{\rm FB}^{(n)}(W_1,W_2)$ be 
a set of all probability distributions 
$\tilde{p}_{L^nX^nY^nZ^n}$ on 
${\cal L}_n$
$\times {\cal X}^n$
$\times {\cal Y}^n$
$\times {\cal Z}^n$
having the form:
\beqno
& &\tilde{p}_{L^nX^nY^nZ^n}(l,x^n,y^n,z^n)
\\
&=&\tilde{p}_{L^n}(l)
\prod_{t=1}^n \left\{
\tilde{p}_{X_t|L_n X^{t-1}Y^{t-1}Z^{t-1}}
(x_t|l,x^{t-1},y^{t-1},z^{t-1}) \right.
\\
& &\qquad \qquad  \times W_1(y_t|x_t)W_2(z_t|y_t)\}.
\eeqno
For simplicity of notation we use the notation $\tilde{p}^{(n)}$ 
for $\tilde{p}_{L_nX^nY^nZ^n}$ $\in {\cal P}_{\rm FB}^{(n)}$
$(W_1,W_2)$. We assume that 
$
\tilde{p}_{U_tX_tY_tZ_t}=\tilde{p}_{L_nX_tY^tZ^t}
$
is a marginal distribution of $\tilde{p}^{(n)}$. 
For $t=1,2,\cdots, n$, we simply write $\tilde{p}_t=$ 
$\tilde{p}_{\empty{U}_tX_tY_tZ_t}$. 
For $\tilde{p}^{(n)}$ $\in {\cal P}_{\rm FB}^{(n)}(W_1,W_2)$ 
and $\tilde{q}^n$ $\in \tilde{\cal Q}^n$, we define  
\beqno
&&
{\Omega}_{\tilde{p}^{(n)}||\tilde{q}^n}^{(\mu,\theta)}(X^nY^nZ^n|L_n)
\\
&&
\defeq \log {\rm E}_{\tilde{p}^{(n)}}
\left[
\prod_{t=1}^n
\frac{W_1^{\theta\mu}(Y_t|X_t)}
     {\tilde{q}^{\theta\mu}_{Y_t|U_t}(Y_t|U_t)}
\frac{\tilde{p}^{\theta}_{Z_t|V_t}(Z_t|V_t)}
{\tilde{q}^{\theta}_{Z_t}(Z_t)}
\right],
\eeqno
where for each $t=1,2,\cdots,n$, the conditional 
probability distribution $\tilde{q}_{Y_t|U_t}$ 
and the probability distribution $\tilde{q}_{Z_t}$
appearing in the definition of 
$
{\Omega}^{(\mu,\theta)}_{\tilde{p}^{(n)}||\tilde{q}^{n}}
(X^nY^nZ^n|L_n)
$
are chosen so that they are induced by the joint distribution 
$\tilde{q}_t=\tilde{q}_{U_tX_tY_tZ_t}\in \tilde{\cal Q}_t$. 
Set 
\beqno
& &
\overline{{\Omega}}_{\rm FB}^{(\mu,\theta)}(W_1,W_2)
\\
&\defeq & 
\sup_{n\geq 1}
\max_{
          { \scs 
        \tilde{p}^{(n)}
          \in {\cal P}_{\rm FB}^{(n)}(W_1,W_2)
          }
     }
\min_{\scs \tilde{q}^n \in \tilde{\cal Q}^n }
\frac{1}{n}{\Omega}_{\tilde{p}^{(n)}||\tilde{q}^n}^{(\mu,\theta)}(X^nY^nZ^n|{L_n}).
\eeqno
Then we have the following proposition.
\begin{pro} 
\label{pro:FbProOne}
For any $\theta >0,\mu>0$, we have 
$$
G_{\rm FB}(R_1,R_2|W_1,W_2)\geq 
\frac{\theta(\mu R_1+R_2)-\overline{\Omega}_{\rm FB}^{(\mu,\theta)}(W_1,W_2)
} {1+\theta(1+\mu)}.
$$
\end{pro}

Proof of this proposition is in Appendix \ref{sub:apdfbproof}.
We shall call $\overline{\Omega}_{\rm FB}^{(\mu,\theta)}(W_1,W_2)$ 
the communication potential. The above corollary implies that 
the analysis of $\overline{\Omega}_{\rm FB}^{(\mu,\theta)}($ $W_1,W_2)$ 
leads to an establishment of a strong converse 
theorem for the degraded BC with feedback. 

\newcommand{\ApdFbProof}{
\subsection{Proof of Proposition \ref{pro:FbProOne}}
\label{sub:apdfbproof}

In this appendix we prove Proposition \ref{pro:FbProOne}. 
We use the following lemma, which is well known as the Cram\`er's bound in 
the large deviation principle.
\begin{lm}
\label{lm:Ohzzzb}
For any real valued random variable $Z$ and any $\theta>0$, 
we have
$$
\Pr\{Z \geq a \}\leq 
\exp
\left[
-\left(
\lambda a -\log {\rm E}[\exp(\theta Z)]
\right) 
\right].
$$
\end{lm}

By Lemmas \ref{lm:FbOhzzzb} and \ref{lm:Ohzzzb}, we have the 
following proposition.
\begin{pro}
\label{pro:FbOhzzp}
For any $\mu,$ $\theta >0$, any $(\tilde{\varphi}^{n},\psi_1^{(n)},\psi_2^{(n)})$  
satisfying  
\beq
\frac{1}{n}\log |{\cal K}_n| \geq R_1,
\frac{1}{n}\log |{\cal L}_n| \geq R_2,
\label{eqn:FXasDD}
\eeq
and any $\tilde{q}^n \in \tilde{\cal Q}^n$, we have 
\beqno	
& & {\rm P}_{\rm c,FB}^{(n)}(\tilde{\varphi}^{n},\psi_1^{(n)},\psi_2^{(n)})
\\
&\leq &3\exp
\left\{
-n\frac{\theta(\mu R_1+R_2)-\frac{1}{n}
{\Omega}_{\tilde{p}^{(n)}||\tilde{q}^n}^{(\mu,\theta)}(X^nY^nZ^n|{L_n})}
{1+\theta(1+\mu)}
\right\}.
\eeqno
\end{pro}

{
{\it Proof:} Under the condition (\ref{eqn:FXasDD}), 
we have the following chain of inequalities:
\beqa
& &
{\rm P}_{\rm c,FB}^{(n)}(\tilde{\varphi}^{n},\psi_1^{(n)},\psi_2^{(n)})
\MLeq{a} \tilde{p}_{{L_n}X^nY^nZ^n} \hugel
\nonumber\\
& &R_1\leq \frac{1}{n}
\sum_{t=1}^n
\log \frac{W_1(Y_t|X_t)}
{\tilde{q}_{Y_t|\empty{U}_t}(Y_t|\empty{U}_t)}+\eta,
\nonumber\\
& & R_2 \leq 
\frac{1}{n}
\sum_{t=1}^n\log
\frac{ \tilde{p}_{Z_t|\empty{V}_t}(Z_t|\empty{V}_t)}
{\tilde{q}_{Z_t}(Z_t)}+\eta
\huger
+2{\rm e}^{-n\eta} 
\nonumber\\ 
&\leq &\tilde{p}_{{L_n}X^nY^nZ^n}\hugel
\mu R_1+R_2-(\mu+1)\eta 
\nonumber\\
& &
\left. \leq \frac{1}{n}
\sum_{t=1}^n
\log \left[
\frac{W_1(Y_t|X_t) \tilde{p}_{Z_t|\empty{V}_t}(Z_t|\empty{V}_t)}
{\tilde{q}^{\mu}_{Y_t| \empty{U}_t}(Y_t|\empty{U}_t)
\tilde{q}^{\mu}_{Z_t}(Z_t)}
\right]\right\}
+2{\rm e}^{-n\eta} 
\nonumber\\
&\MLeq{b} &
\exp\Bigl[n\Bigl\{ -\theta(\mu R_1+R_2)+\theta(\mu+1)\eta 
\nonumber\\
&&\qquad \left.\left. 
+\frac{1}{n}\Omega^{(\mu,\theta)} (X^nY^nZ^n|{L_n})\right\}\right]
+2{\rm e}^{-n\eta}. 
\label{eqn:Fbaaabv}
\eeqa
Step (a) follows from Lemma \ref{lm:FbOhzzzb}.
Step (b) follows from Lemma \ref{lm:Ohzzzb}. 
We choose $\eta$ so that 
\beqa
-\eta&=& -\theta(\mu R_1+R_2)+\theta(\mu+1)\eta
\nonumber\\
& & +\frac{1}{n}\Omega_{\tilde{p}^{(n)}||\tilde{q}^n}^{(\mu,\theta)}(X^nY^nZ^n|{L_n}).
\label{eqn:Fbaaappp}
\eeqa
Solving (\ref{eqn:Fbaaappp}) with respect to $\eta$, we have 
\beqno
\eta=
\frac{
\theta(\mu R_1+R_2)-
\frac{1}{n}{\Omega}_{\tilde{p}^{(n)}||\tilde{q}^n}
^{(\mu,\theta)}(X^nY^nZ^n|{L_n})
}
{1+\theta(1+\mu)}.
\eeqno
For this choice of $\eta$ and (\ref{eqn:Fbaaabv}), we have
\beqno
& &{\rm P}_{\rm c,FB}^{(n)}%
\leq 3{\rm e}^{-n\eta}
\\
&=&3\exp
\left\{
-n\frac{
\theta(\mu R_1+R_2)
-\frac{1}{n}
{\Omega}_{\tilde{p}^{(n)}||\tilde{q}^n}^{(\mu,\theta)}(X^nY^nZ^n|{L_n})
}
{1+\theta(1+\mu)}
\right\},
\eeqno
completing the proof. 
\hfill \IEEEQED
}

{\it Proof of Proposition \ref{pro:FbProOne}}   
By the definitions of $G_{\rm FB}^{(n)}(R_1,$ $R_2|W_1,W_2)$ 
and $\overline{\Omega}_{\rm FB}^{(\mu,\theta)}(W_1,W_2)$
and Proposition \ref{pro:FbOhzzp}, we have 
\beqa
& &G_{\rm FB}^{(n)}(R_1,R_2|W_1,W_2)
\nonumber\\
&\geq & 
\frac{
\theta(\mu R_1+R_2)-
\overline{\Omega}_{\rm FB}^{(\mu,\theta)}(W_1,W_2)
}{1+\theta(1+\mu)}
-\frac{1}{n}\log 3.
\label{eqn:Fbaaaxc}
\eeqa
From (\ref{eqn:Fbaaaxc}), we have Proposition \ref{pro:FbProOne}.
\hfill\IEEEQED
}
The following proposition is a mathematical core 
to prove our main result.
\begin{pro}\label{pro:Fbmainpro}
For $\theta\in (0,1)$, set 
\beq
\lambda = \frac{\theta}{1-\theta} 
\Leftrightarrow \theta=\frac{\lambda}{1+\lambda}. 
\label{eqn:Fbabaddd}
\eeq 
Then, for any $\theta \in (0,1)$, we have 
$$
\overline{\Omega}_{\rm FB}^{(\mu,\theta)}(W_1,W_2)
\leq \frac{1}{1+\lambda}\Omega^{(\mu,\lambda)}(W_1,W_2).
$$
\end{pro}

Proof of this proposition is in Appendix \ref{sub:SSxc}. 
The proof is not so simple. We must introduce 
a new method for the proof. 

\newcommand{\ApdFbArgument}{
%
%
\subsection{
Upper Bound of $\overline{\Omega}_{\rm FB}^{(\mu,\theta)}(W_1,W_2)$
}\label{sub:SSxc}

In this appendix we drive an explicit 
upper bound of $\overline{\Omega}_{\rm FB}^{(\mu,\theta)}(W_1,W_2)$ 
to prove Proposition \ref{pro:Fbmainpro}. 
For each $t=1,2,$ $\cdots,n$, define the function of 
$(u_t,x_t,y_t,z_t)$
$\in \empty{\cal V}_t$
$\times {\cal X}$
$\times {\cal Y}$
$\times {\cal Z}$ 
by 
\beqno
f_{\tilde{p}_t||\tilde{q}_t, \kappa_t}^{(\mu,\lambda)}
(x_t,y_t,z_t|u_t)
&\defeq& 
\frac{W_1^{\theta\mu}(y_t|x_t)
\tilde{p}_{Z_t|V_t}^{\theta}(z_t|v_t)}
{\tilde{q}_{Y_t|U_t}^{\theta \mu}(y_t|u_t)
 \tilde{q}_{Z_t}^{\theta}(z_t)}. 
\eeqno
For each $t=1,2,\cdots,n$, we define the probability distribution
\beqno
& &\tilde{p}_{L_nX^tY^tZ^t}^{(\mu,\theta;\tilde{q}^t,\kappa^t)}
\\
&\defeq& 
\left\{
\tilde{p}_{L_n X^tY^tZ^t}^{(\mu,\theta;\tilde{q}^t,\kappa^t)}(l,x^t,y^t,z^t)
\right\}_{(l,x^t,y^t,z^t)
\in {\cal L}_n \times {\cal X}^t \times {\cal Y}^t 
\times {\cal Z}^t}
\eeqno
by 
\beqno
& &
\tilde{p}_{L_n X^t Y^t Z^t}^{(\mu,\theta;\tilde{q}^t,\kappa^t)}(l,x^t,y^t,z^t) 
\\
&\defeq& \tilde{C}_t^{-1} p_{L_n}(l)
\prod_{i=1}^t \{ \tilde{p}_{X_i|U_iX^{i-1}}
(x_i|u_i,x^{i-1}) 
\\
& &\times W_1(y_i|x_i)W_2(z_i|y_i)
    \empty{f}_{\tilde{p}_i||\tilde{q}_i,\kappa_i}^{(\mu,\theta)}
    (x_i,y_i,z_i|u_i)\},
\eeqno
where
\beqno
\tilde{C}_t 
&\defeq & \sum_{l,x^t,y^t,z^t} 
p_{L_n}(l)\prod_{i=1}^t \{ \tilde{p}_{X_i|U_iX^{i-1}}
(x_i|u_i,x^{i-1}) 
\\
& &\times W_1(y_i|x_i)W_2(z_i|y_i)
    f_{\tilde{p}_i||\tilde{q}_i,\kappa_i}^{(\mu,\theta)}
    (x_i,y_i,z_i|u_i)\}.
\eeqno
are constants for normalization. For each $t=1,2, \cdots, n$, 
set 
\beq 
\tilde{\Phi}_{t,\tilde{q}^t,\kappa^t}^{(\mu,\theta)}\defeq \tilde{C}_t \tilde{C}_{t-1}^{-1},
\label{eqn:Fbdefa}
\eeq
where we define $\tilde{C}_0=1$. 
Then we have the following lemma.
\begin{lm}\label{lm:Fbkeylm}
\beqa 
& &{\Omega}^{(\mu,\theta)}_{\tilde{p}^{(n)}||\tilde{q}^n}
    (X^nY^nZ^n|{L_n})
 =\sum_{t=1}^n \log \tilde{\Phi}_{t,\tilde{q}^t,\kappa^t}^{(\mu,\theta)}.
\label{eqn:Fbdefazzz}  
\eeqa
\end{lm}

{\it Proof}: From (\ref{eqn:Fbdefa}) we have
\beq
\log \tilde{\Phi}_{t,\tilde{q}^t,\kappa^t}^{(\mu,\theta)}
=\log \tilde{C}_t - \log \tilde{C}_{t-1}. 
\label{eqn:Fbaaap}
\eeq
Furthermore, by definition we have 
\beq
{\Omega}_{\tilde{p}^{(n)}||\tilde{q}^n}^{(\mu,\theta)}(X^nY^nZ^n|{L_n})
= \log \tilde{C}_n, \tilde{C}_0=1. 
\label{eqn:Fbaaapq}
\eeq
From (\ref{eqn:Fbaaap}) and (\ref{eqn:Fbaaapq}), 
(\ref{eqn:Fbdefazzz}) is obvious. \hfill \IEEEQED

The following lemma is useful for the computation of 
$\tilde{\Phi}_{t,\tilde{q}^t,\kappa^t}^{(\mu,\theta)}$
for $t=1,2,\cdots,n$.
\begin{lm}\label{lm:Fbaaa}
For each $t=1,2,\cdots,n$, and for any 
$(l,$ $x^t, y^t,z^t)\in {\cal L}_n$ 
$\times {\cal X}^t$
$\times {\cal Y}^t$
$\times {\cal Z}^t$,
we have
\beqa
& &\tilde{p}_{L_n X^t Y^t Z^t}^{(\mu,\theta;\tilde{q}^t,\kappa^t)}
(l,x^t,y^t,z^t)
\nonumber\\
& &=(\tilde{\Phi}_{t,\tilde{q}^t,\kappa^t}^{(\mu,\theta)})^{-1}
\tilde{p}_{L_n X^{t-1} Y^{t-1} Z^{t-1} }
^{(\mu,\theta;\tilde{q}^{t-1}, \kappa^{t-1})}
(l,x^{t-1},y^{t-1}, z^{t-1})
\nonumber\\
& &\quad \times 
\tilde{p}_{X_t|U_tX^{t-1}}
(x_t|\empty{u}_t,x^{t-1})W_1(y_t|x_t)W_2(z_t|y_t)
\nonumber\\
& &\quad \times 
{f}_{\tilde{p}_t||\tilde{q}_t, \kappa_t}^{(\mu,\theta)}(x_t,y_t,z_t|u_t).
\label{eqn:Fbsatt}
\eeqa
Furthermore, we have
\beqa
\tilde{\Phi}_{t,\tilde{q}^t,\kappa^t}^{(\mu,\theta)}
&=& \sum_{l,x^t,y^t, z^t} 
\tilde{p}_{L_n X^{t-1} Y^{t-1} Z^{t-1}}^{(\mu,\theta;\tilde{q}^{t-1},\kappa^{t-1})}
(l,x^{t-1},y^{t-1}, z^{t-1})
\nonumber\\
& &\times 
\tilde{p}_{X_t|U_tX^{t-1}}(x_t|u_t,x^{t-1})
W_1(y_t|x_t)W_2(z_t|y_t)
\nonumber\\
& &\times 
f_{\tilde{p}_t||\tilde{q}_t,\kappa_t}^{(\mu,\theta)}
(x_t,y_t,z_t|u_t)
\nonumber\\
&=&\sum_{u_t,x^{t},y_t, z_t} 
\tilde{p}_{U_tX^{t-1}}^{(\mu,\theta;\tilde{q}^{t-1}, \kappa^{t-1})}
(u_t,x^{t-1})
\nonumber\\
& & \times \tilde{p}_{X_t|U_tX^{t-1}}
(x_t|u_t,x^{t-1})W_1(y_t|x_t)W_2(z_t|y_t)
\nonumber\\
& & \times f_{\tilde{p}_t||\tilde{q}_t, \kappa_t}^{(\mu,\theta)}
(x_t,y_t,z_t|u_t).
\label{eqn:Fbsattb}
\eeqa
\end{lm}
}
\newcommand{\ApdcFb}{

{\it Proof:} 
By the definition of 
$\tilde{p}_{{L_n}X^tY^{t}Z^{t}}^{(\mu,\theta;\tilde{q}^t,\kappa^t)}$ 
$(l,$ $x^t,y^t,z^t)$, 
$t=1,2,\cdots,n$, we have 
\beqa
\hspace*{-4mm}
& &\tilde{p}_{L_nX^tY^{t}Z^{t}}
^{(\mu,\theta;\tilde{q}^t,\kappa^t)}(l,x^t,y^{t},z^{t})
\nonumber\\
\hspace*{-4mm}
&=&\tilde{C}_t^{-1}
p_{L_n}(l)
\prod_{i=1}^t\{
\tilde{p}_{X_i|U_iX^{i-1}}(x_i|u_i,x^{i-1}) 
\nonumber\\
& &
\times W_1(y_i|x_i)W_2(z_i|y_i)
f_{\tilde{p}_i||\tilde{q}_i,\kappa_i}^{(\mu,\theta)}
(x_i,y_i,z_i|u_i)\}.
\label{eqn:Fbazaq}
\eeqa 
Then we have the following chain of equalities:
\beqa 
& &\tilde{p}_{L_nX^tY^tZ^t}^{(\mu,\theta;\tilde{q}^t,\kappa^t)}(l,x^t,y^t,z^t)
\nonumber\\
&\MEq{a}&
\tilde{C}_t^{-1} 
p_{L_n}(l)
\prod_{i=1}^t \{ 
\tilde{p}_{X_i|U_i X^{i-1}}(x_i|u_i,x^{i-1})
\nonumber\\
& &
\times 
W_1(y_i|x_i)W_2(z_i|y_i)
f_{\tilde{p}_i||\tilde{q}_i,\kappa_i}^{(\mu,\theta)}
(x_i,y_i,z_i|u_i)\}
\nonumber\\
&=&\tilde{C}_t^{-1}p_{L_n}(l)
\prod_{i=1}^{t-1}\{
\tilde{p}_{X_i|U_iX^{i-1} }(x_i|u_i,x^{i-1})
\nonumber\\
&&\times
W_1(y_i|x_i)W_2(z_i|y_i)
f_{\tilde{p}_i||\tilde{q}_i,\kappa_i}^{(\mu,\theta)}
(x_i,y_i,z_i|u_i)\}
\nonumber\\
& &\times 
\tilde{p}_{X_t|U_tX^{t-1}}(x_t|u_t,x^{t-1})
W_1(y_t|x_t)W_2(z_t|y_t) 
\nonumber\\
& &\times 
{f}_{\tilde{p}_t||\tilde{q}_t, \kappa_t}^{(\mu,\theta)}(x_t,y_t,z_t|u_t)
\nonumber\\
&\MEq{b}&
\tilde{C}_t^{-1}\tilde{C}_{t-1}
\tilde{p}_{{L_n}X^{t-1}Y^{t-1}Z^{t-1}}^{(\mu,\theta;\tilde{q}^{t-1}, \kappa^{t-1})}
(l,x^{t-1},y^{t-1}, z^{t-1})
\nonumber\\
& &\times 
\tilde{p}_{X_t|U_tX^{t-1}}(x_t|u_t,x^{t-1})
W_1(y_t|x_t)W_2(z_t|y_t) 
\nonumber\\
& & 
\times 
f_{\tilde{p}_t||\tilde{q}_t, \kappa_t}^{(\mu,\theta)}(x_t,y_t,z_t|u_t)
\nonumber\\
&=&(\tilde{\Phi}_{t,\tilde{q}^t,\kappa^t}^{(\mu,\theta)})^{-1}
\tilde{p}_{{L_n}X^{t-1}Y^{t-1}Z^{t-1}}^{(\mu,\theta;\tilde{q}^{t-1},\kappa^{t-1})}
(l,x^{t-1},y^{t-1},z^{t-1})
\nonumber\\
& &\times 
\tilde{p}_{X_t|U_tX^{t-1}}(x_t|u_t,x^{t-1})
W_1(y_t|x_t)W_2(z_t|y_t) 
\nonumber\\
& &
\times {f}_{\tilde{p}_t||\tilde{q}_t, \kappa_t}^{(\mu,\theta)}(x_t,y_t,z_t|u_t).
\label{eqn:Fbdaaaq}
\eeqa
Steps (a) and (b) follow from (\ref{eqn:Fbazaq}). 
From (\ref{eqn:Fbdaaaq}), we have 
\beqa 
& &\tilde{\Phi}_{t,\tilde{q}^t,\kappa^t}^{(\mu,\theta)}
\tilde{p}_{{L_n}X^tY^tZ^t}^{(\mu,\theta ;\tilde{q}^t,\kappa^t)}(l,x^t,y^t,z^t)
\label{eqn:Fbdaxx}\\
&=&\tilde{p}_{{L_n}X^{t-1}Y^{t-1}Z^{t-1}}
^{(\mu,\theta;\tilde{q}^{t-1}, \kappa^{t-1})}
(l,x^{t-1},y^{t-1},z^{t-1})
\nonumber\\
& &\times 
\tilde{p}_{X_t|U_tX^{t-1}}(x_t|u_t,x^{t-1})
W_1(y_t|x_t)W_2(z_t|y_t) 
\nonumber\\
& &\times
f_{\tilde{p}_t||\tilde{q}_t,\kappa_t}^{(\mu,\theta)}(x_t,y_t,z_t|u_t).
\label{eqn:Fbdaaxx}
\eeqa
Taking summations of (\ref{eqn:Fbdaxx}) and 
(\ref{eqn:Fbdaaxx}) with respect to $l,x^t,$ $y^t,$ $z^t$, 
we obtain 
\beqa
& &\tilde{\Phi}_{t,\tilde{q}^t,\kappa^t}^{(\mu,\theta)}
\nonumber\\
&=& \sum_{l,x^t,y^t,z^t}
\tilde{p}_{{L_n}X^{t-1}Y^{t-1}Z^{t-1}}^{(\mu,\theta;\tilde{q}^{t-1}, \kappa^{t-1})}
(l,x^{t-1},y^{t-1},z^{t-1})
\nonumber\\
& & \times 
  \tilde{p}_{X_t|U_t,X^{t-1}}
  (x_t|u_t,x^{t-1})W_1(y_t|x_t)W_2(z_t|y_t)
\nonumber\\
& & \times 
f_{\tilde{p}_t||\tilde{q}_t,\kappa_t}^{(\mu,\theta)}(x_t,y_t,z_t|u_t)
\nonumber\\
&=&\sum_{u_t,x^{t},y_t, z_t} 
\tilde{p}_{U_tX^{t-1}}^{(\mu,\theta;\tilde{q}^{t-1}, \kappa^{t-1})}
(\empty{u}_t,x^{t-1})
\nonumber\\
& & \times \tilde{p}_{X_t|U_t,X^{t-1}}
(x_t|\empty{u}_t,x^{t-1})W_1(y_t|x_t)W_2(z_t|y_t)
\nonumber\\
& & \times f_{\tilde{p}_t||\tilde{q}_t,\kappa_t}^{(\mu,\theta)}
(x_t,y_t,z_t|u_t),
\nonumber
\eeqa
completing the proof.
\hfill \IEEEQED
}
\newcommand{\ApdFbArgumentB}{

We set 
\beqno
& &\tilde{p}_{U_tX_t}^{(\mu,\theta;\tilde{q}^{t-1}, \kappa^{t-1})}(u_t,x_t)
\\
&=&\sum_{x^{t-1}}
\tilde{p}_{U_tX^{t-1}}^{(\mu,\theta;\tilde{q}^{t-1}, \kappa^{t-1})}(u_t,x^{t-1})
\tilde{p}_{X_t|U_tX^{t-1}}(x_t|u_t,x^{t-1}).
\eeqno
Then by (\ref{eqn:Fbsattb}) in Lemma \ref{lm:Fbaaa} and the definition 
of $f^{(\mu,\theta)}_{\tilde{p}_t||\tilde{q}_t,\kappa_t}$
$(x_t$$,y_t,$$z_t$$|u_t)$, we have 
\beqa
& &
\tilde{\Phi}_{t,\tilde{q}^t,\kappa^t}^{(\mu,\theta)}
\nonumber\\
&=&
\sum_{\empty{u}_t, x_t, y_t, z_t}
\tilde{p}_{U_t X_t}^{(\mu,\theta;\tilde{q}^{t-1}, \kappa^{t-1})}(u_t,x_t)
W_1(y_t|x_t)W_2(z_t|y_t)
\nonumber\\
& &\quad\times 
\frac{W_1^{\theta\mu}(y_t|x_t)\tilde{p}_{Z_t|V_t}^{\theta}(z_t|v_t)}
{\tilde{q}_{Y_t|U_t}^{\theta\mu}(y_t|u_t)\tilde{q}_{Z_t}^{\theta}(z_t)}.
\label{eqn:Fbaasor}
\eeqa
}
\newcommand{\ApdFbArgumentC}{
%
%
Proof of Proposition \ref{pro:Fbmainpro} is as follows.
 
{\it Proof of Proposition \ref{pro:Fbmainpro}:} 
Set
\beqno
&&\tilde{\cal P}_n(W_1,W_2)
\defeq 
\{\tilde{q}=\tilde{q}_{UXYZ}:
\pa {\cal U} \pa \leq \pa {\cal L}_n\pa 
\pa {\cal Y}\pa^{n-1} \pa {\cal Z}\pa^{n-1},
\vSpa\\
&&\qquad 
\tilde{q}_{Y|X}=W_1,\tilde{q}_{Z|Y}=W_2,  
U \markov X\markov Y \markov Z \},
\\
& &\tilde{\Omega}_n^{(\mu,\lambda)}(W_1,W_2)
\defeq \max_{\tilde{q} \in \tilde{\cal P}_n(W_1,W_2)}
\log \Omega_{\tilde{q}}^{(\mu,\lambda)}({XYZ|U}).
\eeqno
We choose $\tilde{q}_t=$$\tilde{q}_{U_t}$${}_{X_t}$${}_{Y_t}$${}_{Z_t}$ 
so that
\beqno
& &\tilde{q}_{U_tX_tY_tZ_t}(u_t,x_t,y_t,z_t)
\\
&=&\tilde{p}_{U_tX_t}^{(\mu,\theta;\tilde{q}^{t-1}, \kappa^{t-1})} (u_t,x_t)
W_1(y_1|x_t)W_2(z_t|y_t).
\eeqno
It is obvious that $\tilde{q}_t\in\tilde{\cal P}_n(W_1,W_2)$ for 
$t=1,2, \cdots, n$. 
By (\ref{eqn:Fbaasor}) and the above choice of $\tilde{q}_t$, we have 
\beqa
& &\tilde{\Phi}_{t,\tilde{q}^t,\kappa^t}^{(\mu,\theta)}
\nonumber\\
&=&\sum_{u_t,x^t,y_t,z_t} 
\tilde{q}_{U_t}(u_t)\tilde{q}_{X_t|U_t}(x_t|u_t)W_1(y_t|x_t)W_2(z_t|y_t).
\nonumber\\
& &\times 
\left\{
\frac{W_1^{\mu}(y_t|x_t)}{\tilde{q}^{\mu}_{Y_t|U_t}(y_t|u_t)}
\frac{\tilde{p}_{Z_t|V_t}(z_t|v_t)}{\tilde{q}_{Z_t}(z_t)}
\right\}^\theta
\nonumber\\
&=&
{\rm E}_{\tilde{q}_t}
\left[
\left\{
\frac{W_1^{\mu}(Y_t|X_t)}{\tilde{q}^{\mu}_{Y_t|U_t}(Y_t|U_t)}
\frac{\tilde{p}_{Z_t|V_t}(Z_t|V_t)}{\tilde{q}_{Z_t}(V_t)}
\right\}^{\theta}
\right]
\nonumber\\
&=&
{\rm E}_{\tilde{q}_t}
\left[
\left\{
\frac{W_1^{\mu}(Y_t|X_t)}{\tilde{q}^{\mu}_{Y_t|U_t}(Y_t|U_t)}
\frac{\tilde{q}_{Z_t|U_t}(Z_t|U_t)}{\tilde{q}_{Z_t}(Z_t)}
\frac{\tilde{p}_{Z_t|V_t}(Z_t|V_t)}
     {\tilde{q}_{Z_t|U_t}(Z_t|U_t)}
\right\}^{\theta}
\right]
\nonumber\\
&\MLeq{a}&
\left(
{\rm E}_{\tilde{q}_t}
\left[
\left\{
\frac{W_1^{\mu}(Y_t|X_t)}{\tilde{q}^{\mu}_{Y_t|U_t}(Y_t|U_t)}
\frac{\tilde{q}_{Z_t|U_t}(Z_t|U_t) }{\tilde{q}_{Z_t}(Z_t)}
\right\}^{\frac{\theta}{1-\theta}}
\right]
\right)^{1-\theta}
\nonumber\\
& &\times \left(
{\rm E}_{\tilde{q}_t}
\left\{
\frac
{\tilde{p}_{Z_t|\empty{V}_t}(Z_t|\empty{V}_t)}
{\tilde{q}_{Z_t|\empty{U}_t}(Z_t|\empty{U}_t)}
\right\}\right)^{\theta}
\nonumber\\
&=&
\exp\left\{(1-\theta)
   {\Omega}^{(\mu,\frac{\theta}{1-\theta})}_{\tilde{q}_t}
   (X_tY_tZ_t|\empty{U}_t)
\right\}
\nonumber\\
&\MEq{b}&
\exp\left\{\frac{1}{1+\lambda}{\Omega}^{(\mu,\lambda)}_{\tilde{q}_t}
   (X_tY_tZ_t|U_t)
\right\}
\nonumber\\
&\MLeq{c}&
\exp\left\{\frac{1}{1+\lambda}
   \tilde{\Omega}_n^{(\mu,\lambda)}(W_1,W_2)
\right\}
\nonumber\\
&
\MEq{d}
&
\exp\left\{
\frac{1}{1+\lambda}
   {\Omega}^{(\mu,\lambda)}(W_1,W_2)
\right\}.
\label{eqn:Fbsssto} 
\eeqa
Step (a) follows from H\"older's inequality. 
Step (b) follows from (\ref{eqn:Fbabaddd}). 
Step (c) follows from $\tilde{q}_t \in \tilde{\cal P}_n(W_1,W_2)$ 
and the definition of $\tilde{\Omega}_n^{(\mu,\lambda)}$$(W_1,W_2)$. 
Step (d) follows from Lemma \ref{lm:FbCardLm} 
in Appendix \ref{sub:CardBound}. To prove this lemma 
we bound the cardinality $|{\cal V}|$ appearing in  
the definition of $\tilde{\Omega}_n^{(\mu,\lambda)}(W_1,W_2)$ 
to show that the bound $|{\cal U}|\leq |{\cal X}|$ 
is sufficient to describe 
$\tilde{\Omega}_n^{(\mu,\lambda)}(W_1,W_2)$.
Hence we have the following: 
\beqa
& &\min_{\scs \tilde{q}^{n}\in \tilde{\cal Q}^n}
\frac{1}{n}\Omega^{(\mu,\theta)}_{\tilde{p}^{(n)}||\tilde{q}^{n}}(X^nY^nZ^n|{L_n})
\nonumber\\
&\leq &\frac{1}{n}{\Omega}_{\tilde{p}^{(n)}||\tilde{q}^n}^{(\mu,\theta)}(X^nY^nZ^n|L_n)
\MEq{a}\frac{1}{n}\sum_{t=1}^n \log 
\tilde{\Phi}_{t, \tilde{q}^t,\kappa^t}^{(\mu,\theta)}
\nonumber\\
&\MLeq{b}& \frac{1}{1+\lambda}
{\Omega}^{(\mu,\lambda)}(W_1,W_2).
\qquad \label{eqn:FaQ1}
\eeqa
Step (a) follows from (\ref{eqn:Fbdefazzz}) in Lemma \ref{lm:Fbkeylm}. 
Step (b) follows from (\ref{eqn:Fbsssto}).
Since (\ref{eqn:FaQ1}) holds for any ${n\geq 1}$ 
and any $\tilde{p}^{(n)}\in {\cal P}_{\rm FB}^{(n)}$ $(W_1,W_2)$, 
we have  
$$
\overline{\Omega}_{\rm FB}^{(\mu,\theta)}(W_1,W_2)
\leq 
\frac{1}{1+\gamma}
{\Omega}^{(\mu,\lambda)}(W_1,W_2).
$$
Thus, Proposition \ref{pro:Fbmainpro} is proved.
\hfill \IEEEQED
}

{\it Proof of Theorem \ref{Th:Fbmain}: }
For $\theta \in (0,1)$, set 
\beq
\lambda=\frac{\theta}{1-\theta} 
\Leftrightarrow \theta=\frac{\lambda}{1+\lambda}. 
\label{eqn:Fbabadd}
\eeq
Then we have the following:
\beqa
& &G_{\rm FB}(R_1,R_2|W_1,W_2)
\nonumber\\
&\MGeq{a}& 
\frac{
\theta(\mu R_1 + R_2)
-\overline{\Omega}^{(\mu,\theta)}_{\rm FB}(W_1,W_2)
}
{1+\theta(1+\mu)}
\nonumber\\
&\MGeq{b}& 
\frac{\frac{\lambda}{1+\lambda}(\mu R_1 + R_2)
-\frac{1}{1+\lambda}{\Omega}^{(\mu,\lambda)}(W_1,W_2)
}
{1+\frac{\lambda}{1+\lambda}(1+\mu)}
\nonumber\\
&=&
\frac{\lambda(\mu R_1 + R_2)-\Omega^{(\mu,\lambda)}(W_1,W_2)
}
{1+\lambda+\lambda(1+\mu)}
\nonumber\\
&=&{F}^{(\mu,\lambda)}(\mu R_1+R_2|W_1,W_2). 
\label{eqn:akka}
\eeqa
Step (a) follows from Proposition \ref{pro:FbProOne}. Step (b) follows from 
Proposition \ref{pro:Fbmainpro} and (\ref{eqn:Fbabadd}).
Since (\ref{eqn:akka}) holds for any positive $\lambda$ and $\mu$, we have
$$
G_{\rm FB}(R_1,R_2|W_1,W_2)\geq F(R_1,R_2|W_1,W_2). 
$$ 
Thus (\ref{eqn:FbmainIeq}) in Theorem \ref{Th:Fbmain} is proved. 
\hfill\IEEEQED
%
%
\newcommand{\ApdaAAAbFb}{
\subsection{Cardinality Bound of Auxilary Random Variables}
\label{sub:CardBound}
We prove the following lemma.
\begin{lm} \label{lm:FbCardLm} 
For each integer $n \geq 2$, we have 
\beqno
&      &\tilde{\Omega}_n^{(\mu,\lambda)}(W_1,W_2)
\\
&\defeq& \max_{\scs q=q_{UXYZ}: U\leftrightarrow X \leftrightarrow Y \leftrightarrow Z,
        \atop{\scs
           \atop{\scs
           q_{Y|X}=W_1,q_{Z|Y}=W_2,
            \atop{\scs  
            |{\cal U}|\leq |{\cal L}_n||{\cal Y}|^{n-1}|{\cal Z}|^{n-1}  
     }}}}
\Omega^{(\mu,\lambda)}_{q}(XYZ|U)
\\
&=& \max_{\scs q=q_{UXYZ}: U\leftrightarrow X \leftrightarrow Y \leftrightarrow Z,
        \atop{\scs
           \atop{\scs
           q_{Y|X}=W_1,q_{Z|Y}=W_2,
            \atop{\scs  
            |{\cal U}|\leq |{\cal X}|
     }}}}
\Omega^{(\mu,\lambda)}_{q}(XYZ|U)
\\&=&\Omega^{(\mu,\lambda)}(W_1,W_2).
\eeqno
\end{lm}

{\it Proof:} We bound  the cardinality $|{\cal U}|$ of ${U}$ 
to show that the bound 
$|{\cal U}| \leq |{\cal X}|$
is sufficient to describe 
$\tilde{\Omega}_n^{(\mu,\lambda)}$ 
$(W_1,W_2)$. 
Observe that 
\beqa
\hspace*{-5mm}& &q_{{X}}(x)
=\sum_{u\in {\cal U}}q_U(u)
q_{{X}|U}(x|u),
\label{eqn:Fbasdf}
\\
\hspace*{-5mm}
& &\Lambda_{q}^{(\mu, \lambda)}(XYZ|U)
=\sum_{u\in {\cal U}}q_U(u)
\zeta^{(\mu, \theta)}(q_{{X}|U}(\cdot|u)),
\label{eqn:Fbaqqqa}
\eeqa
where 
\beqno
& &\zeta^{(\mu, \lambda)}(q_{{X}|U}(\cdot|u))
\\
&\defeq & \sum_{(x,y,z)\in{\cal X}\times{\cal Y}\times{\cal Z}}
q_{{X}|U}(x|u)W_1(y|x)W_2(z|y)
\\
& &\times \exp\left\{\lambda \omega^{(\mu)}_{q}(x,y,z|u)\right\}
\eeqno
are continuous functions of $q_{{X}|U}(\cdot|u)$ .
Then by the support lemma, 
$$
|{\cal U}| \leq |{\cal X}|-1 +1= |{\cal X}| 
$$
is sufficient to express $|{\cal X}|-1$ 
values of (\ref{eqn:Fbasdf}) 
and one value of (\ref{eqn:Fbaqqqa}). 
\hfill \IEEEQED
}
%
%
%

\section*{\empty}
\appendix

\ApdaAAAbFb

\ApdaFb

\ApdLemmaA

\ApdFbProof
\ApdFbArgument
\ApdcFb
\ApdFbArgumentB
\ApdFbArgumentC

\end{document}